\documentclass[12pt]{article}

\usepackage{epsfig}

\begin{document}

\setcounter{page}{1}
\begin{flushright}
DFTUZ 96/1
\end{flushright}
\vskip 10pt
\vskip 10pt

\centerline{\Large{\bf{A multisite microcanonical updating method}}}

\bigskip

\begin{center}
A.~Cruz$^a$, L.~A.~Fern\'andez$^b$,\\
D.~I\~niguez$^a$, and A.~Taranc\'on$^a$\\
\bigskip

{\it
a)  Departamento de F\'{\i}sica Te\'orica, Facultad de Ciencias,\\
Universidad de Zaragoza, 50009 Zaragoza, Spain \\
b)  Departamento de F\'{\i}sica Te\'orica I, Facultad de Ciencias
F\'{\i}sicas,\\
Universidad Complutense de Madrid, 28040 Madrid, Spain \\
}
\bigskip
\today

\end{center}

\bigskip
\begin{abstract}
We have made a study of several update algorithms using the XY
model. We find that sequential local overrelaxation is not ergodic at
the scale of typical Monte Carlo simulation time.  We have introduced
a new multisite microcanonical update method, which yields results
compatible with those of random overrelaxation and the microcanonical
demon algorithm, which are very much slower, all being incompatible
with the sequential overrelaxation results.

\end{abstract}

\newpage

Microcanonical local algorithms are used often in numerical
simulations of spin systems or lattice gauge theories.
Sometimes they are used in combination with canonical algorithms
in order to accelerate decorrelation in Monte Carlo time. Some other
times one is interested in a pure microcanonical simulation for physical 
reasons~\cite{FIRSTORDER,RG,MFA}. 
We do not consider nonlocal microcanonical methods which are based on
molecular dynamics and are used specially for studying dynamical
fermions\cite{MOLECULARDYNAMICS}.

Possibly, the most popular microcanonical algorithm is Local
Microcanonical Overrelaxation (LMO) \cite {CREUTZ1}. The algorithm,
simple and fast, has a dynamical exponent $z=1$ when the update is
done sequentially, and $z=2$ when it is done randomly \cite {SOKAL},
the difference being due to a wave effect occurring in the former
case, which causes the change made in the update of a variable to
propagate to the next one, the updated configuration being farther
from the original one than in the case consecutive updates had been
made independently. Yet, no proof exists of the algorithm being
ergodic. Its determinism, particularly in the case of sequential
update, might induce one to think that it is not ergodic. Indeed, we
shall present evidence of the lack of ergodicity of the sequential LMO
method.

We shall introduce here the Multisite Microcanonical method (MM), a
new microcanonical updating algorithm which exchanges energy among
different regions of the lattice at each update. It has two features
which are absent from the LMO: it is non-local and
non-deterministic. They contribute to diminishing Monte Carlo
correlation time, and avoid the biases that plague LMO. 

Apart from the mentioned LMO and MM, we have also done simulations using the
Microcanonical Demon Algorithm, (MDA), introduced by Creutz \cite {CREUTZ2},
in order to have one further point of reference.

We have chosen a simple model for our simulations, the two dimensional XY
model. We remark that the method could be easily generalized to O($N$)
models in arbitrary dimensions, or even to abelian and SU(2) gauge
theories.

In what follows, we shall recall the two dimensional XY model,
introducing the observables we have measured, and study the mentioned
algorithms, LMO, MDA and MM.

In a square lattice of side $L$, with periodic boundary conditions, we define
on each site a variable $\phi(n) \in U(1)$, the action
\begin{equation}
S=\sum_{n,\mu} {\rm Re \hskip2truemm} \phi(n)\phi^*(n+\mu)
\end{equation}
and the density of states with energy E
\begin{equation}
N(E)={\cal Z}^{-1}\int\prod_n d\phi(n) \delta(S(\phi)-E).
\end{equation}
${\cal Z}$ being the partition function.

The presented results have been obtained in a $64^2$ lattice, with a
normalized energy value $E = 0.4453$, which corresponds to
$\beta\sim 1.0$.  The data plotted in figures have been obtained with
$45,000$ configurations (after thermalization).  Where the results were
not conclusive we have added $500,000$ MC iterations.  Most runs have been
started from two fixed (thermalized) configurations.  Errors have been
calculated using the jackknife method.

Among the measured observables, the energy, the process being
microcanonical, should remain constant. We have measured it as a check
of the different methods, observing small changes due to rounding
errors in the microcanonical methods, and larger oscillations in the
MDA case, because in the latter only the total energy (system plus demon)
remains constant.

We have measured the total magnetization, defined as 
$M=\frac{1}{L^2}\sum_{n}\phi(n)$. $M$ being a complex number, we have measured 
its
modulus and phase, which, although it can be chosen arbitrarily by a global
gauge change, \hbox{$\phi(n)\to U\phi(n)$}, 
$\forall n,\ U \in {\rm U}(1)$,
can give us a hint on how the system is evolving.

The most useful observables for our study have turned out to be the spatial
correlations. In particular, we have measured the correlation between parallel
lines at distances from $0$ to $L/2$, i.e., defining
\begin{equation}
\omega(x)=\frac{1}{L}\sum_{y}\phi(x,y) \qquad
\omega(y)=\frac{1}{L}\sum_{x}\phi(x,y)
\end{equation}
(where the notation
 $\phi(x,y)$ has been used, instead of $\phi(n)$), 
the correlation at a distance $r$ is written
\begin{equation}
\eta(r)=\frac{1}{2} \langle \omega(x) \omega^*(x+r) +
\omega(y) \omega^*(y+r) \rangle.
\end{equation}
We have studied separately the real and imaginary parts of $\eta(r)$. From
the real part, correlation lengths can be extracted. 

The imaginary part should be zero at any $r$, due to the symmetry of
the action under the transformation $\phi(n)\to\phi^*(n), \forall n$. 
We shall see that the imaginary part is not zero
in the LMO method. This implies that the system does not
evolve uniformly in the whole phase-space, so breaking the fundamental
assumption of the microcanonical ensemble: at any time, all the states
of the system with the given energy have equal probability.

Let us consider first the LMO algorithm. 
Writing $\phi=e^{i\theta(n)}$ and defining 
\begin{equation}
\rho(n_0)e^{i\alpha(n_0)}=\sum_{\pm \mu}e^{-i\theta(n_o+\mu)},
\label{POLAR}
\end{equation}
the action associated to the site $n_0$ reads
\begin{equation}
S(n_0)= {\rm Re \hskip2truemm} (\phi(n_0)\sum_{\pm\mu}\phi^*(n_0+\mu))
= \rho(n_0)\cos (\theta(n_0)+\alpha(n_0)).
\end{equation}
One LMO step consists on picking a site $n_0$ and updating $\theta(n_0)$ to a 
new value 
\begin{equation}
\theta^{\prime}(n_0)=-\theta(n_0)-2\alpha(n_0)
\end{equation}
which, due to the parity of the $\cos$ function, does not alter the energy.

The step satisfies detailed balance in a deterministic way: the
probability of going from one state to the other is $1$ in both senses. 
But a simulation takes many of these steps,
and the order in which they are taken is also important. If we
question the relationship between two states separated by $N$ such
steps, taking the intermediate sites randomly satisfies detailed
balance, but taking them sequentially does not. The looser condition
of balance \cite {SOKAL,PARISI} is satisfied, though.  The
latter, together with ergodicity, is sufficient for the algorithm to
reproduce the desired distribution after a reasonable number of
sweeps.  However, the determinacy of each single step makes one question
the ability of the method to satisfy ergodicity. Running a sweep
forward and backwards would leave the configuration unchanged. Also, a
sequential overrelaxation sweep on a one-dimensional XY model would
have as only effect to transport a little energy packet along the
chain, leaving the rest of the chain undisturbed.

Proving ergodicity is, in general, difficult, but disproving it may be
easier. A proof of lack of ergodicity is the
occurrence of results which, in the limit of infinite statistics,
depend on the initial state. Also, finding a conserved quantity which
should not be conserved means that the algorithm runs in a certain
region of phase space, characterized by a certain value of the
conserved quantity, without being able to leave it.

\begin{figure}
\begin{center}
\epsfig{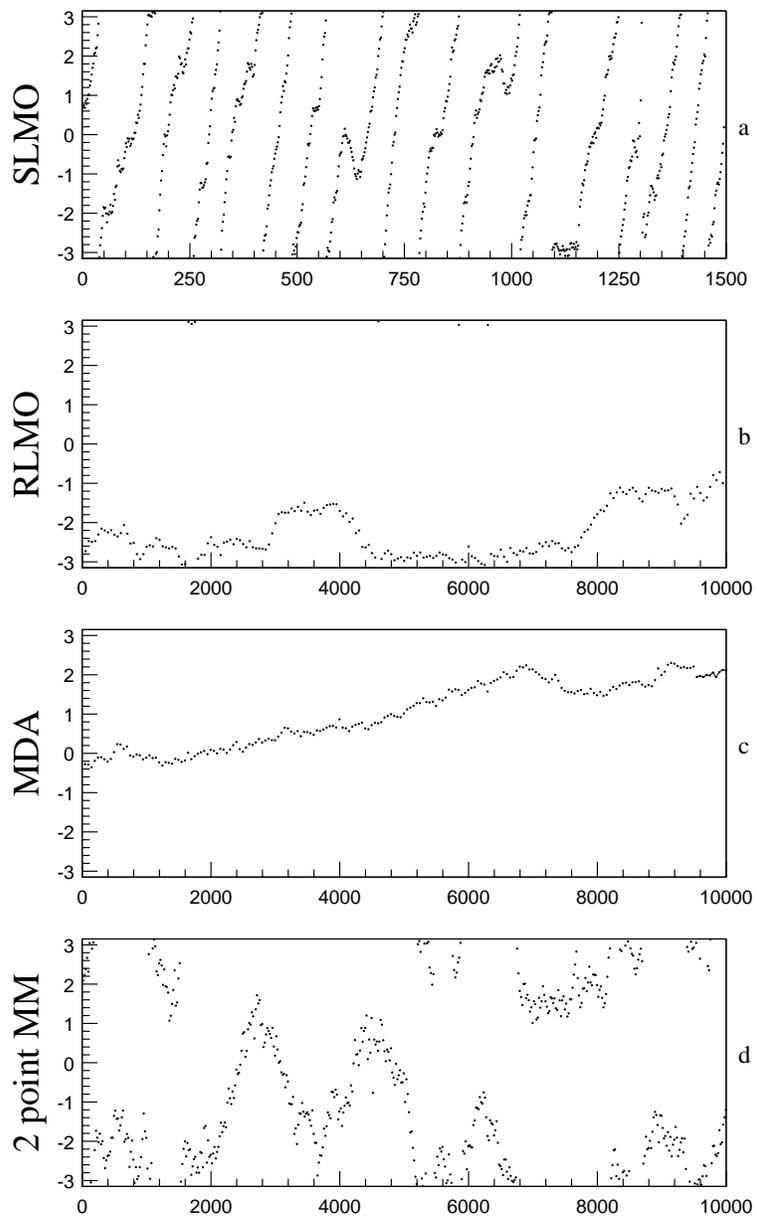}
\caption{Evolution in Monte Carlo time of the phase of the magnetization
for several algorithms. Only a sample of each run is shown.}
\end{center}
\end{figure}

Let us consider sequential LMO, SLMO. Fig.~1a shows that the phase
runs over the interval $(-\pi,\pi]$ at a practically constant rate (we
shall call that property {\em helicity}). The sign and modulus of the
helicity depend on the initial configuration, and they are kept
through the simulation with small fluctuations.  We have consistently
used two fixed configurations, characterized by their respective large
and small value of the helicity, as initial configurations in all our
simulations, in order to check the possible dependence of the results
on the initial state.  Making the necessary allowances for the proviso
made above on the meaning of the phase, the origin of the rotation
described being a global rotation of the system would be of no
importance, but problems might arise if the rotations were local.

The imaginary part of the spatial correlation defined above,
Im$(\eta(r))$, should average statistically to zero, the quantity being
invariant under a global rotation, but sensitive to local
ones. Fig.~2a shows that SLMO yields values for Im$(\eta(r))$ clearly
incompatible with zero within the statistical errors, statistics being
sufficiently high. That, together with the fact that other algorithms
do yield results fully compatible with zero, confirms our suspicion that
SLMO does not run over the phase space properly.

As Im$(\eta(r))$ is not either a very usual observable, we have
considered too Re$(\eta(r))$.
Fig.~3a shows that SLMO yields results incompatible with
each other and dependent on the helicity of the initial configuration.
In order to reinforce this conclusion we have increased the statistics
with $500,000$ more iterations, with results confirming the incompatibility.

\begin{figure}[ht]
\begin{center}
\epsfig{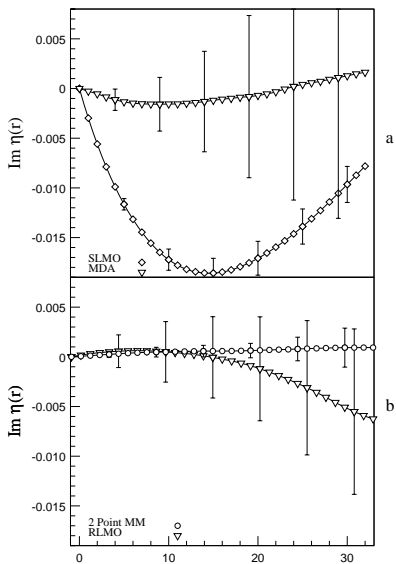}
\caption{Imaginary part of the spatial correlation evaluated by several 
algorithms (with $50,000$ iterations in each case).}
\end{center}
\end{figure}

\begin{figure}[ht]
\begin{center}
\epsfig{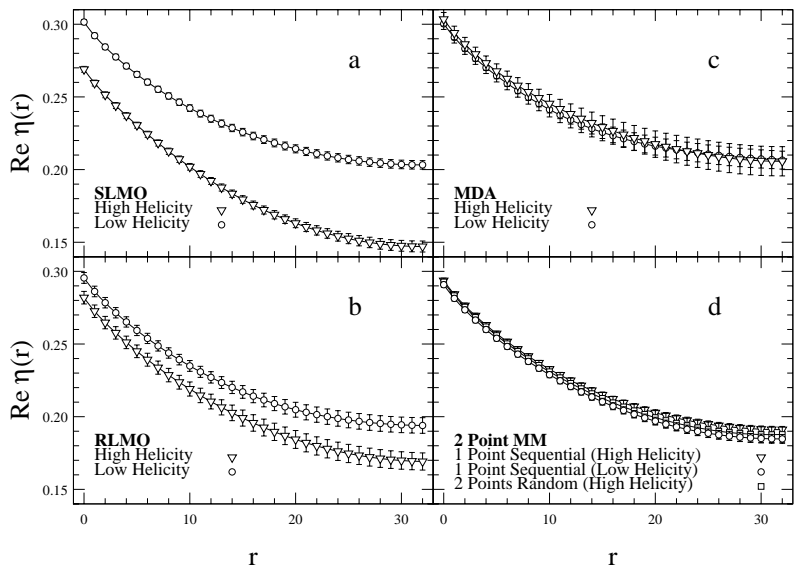}
\caption{Real part of the spatial correlation evaluated by several 
algorithms (with $50,000$ iterations in each case).}
\end{center}
\end{figure}

Randomly ordered LMO (RLMO) does not conserve helicity. In that case,
the phase of $M$, shown in Fig.~1b, evolves very slowly in a
nonsystematic way. Fig.~2b shows that Im$(\eta(r))$ is compatible with
zero. However, Fig.~3b shows that the real part is also 
dependent on the initial configuration,
although the values are less different from each other than in the
SLMO case, and the errors, also for $\vert M \vert$, are much bigger,
and so are the correlation times. Clearly, it would be wrong to infer
from the error size that SLMO is better than RLMO, as we know by now
that the former runs only over a small region of phase space. 

Next, we have run also up to $500,000$ iterations for each
initial configuration (high and low helicity) and in this case the
results for Re$(\eta(r))$ have changed dramatically with respect to those
obtained with $50,000$ iterations. Now the results for the two different
starts become fully compatible (and compatible also with those obtained 
by other methods shown later). Therefore RLMO behaves ergodically, but with
a Monte Carlo time scale very large compared with the scale of the other 
algorithms shown later.

It must be said, in discharge of SLMO, that despite its lack of
ergodicity, the fact that it decorrelates efficiently, in the sense
that it produces configurations which, although all belong to a
restricted region of phase space, are fairly far from each other, can,
as has been widely shown, make it very useful when combined with other
algorithms, like Metropolis, which reinstate ergodicity. 

Let us consider next the Microcanonical Demon Algorithm, in which an
extra degree of freedom, the demon, travels through the system,
transferring energy from one point to another. The total energy of the
system plus the demon is kept constant, which makes the algorithm to
be not strictly microcanonical from the point of view of the system
alone. The algorithm works as follows: a small amount of absolute
energy is assigned initially to the demon, say, 1. When updating a
spin, a new random tentative value is chosen, and the new system energy is
computed. In case it increases, the change is accepted, and the excess
energy is given to the demon. If the energy decreases, the change is
accepted only if the demon energy is sufficient to provide the
difference, the demon energy being always kept non-negative.

As only one demon is being used (more demons can be included in the
formalism) and its initial energy is small, the system evolves in a
practically microcanonical way.  Fig.~1c shows the evolution of the
phase of $M$, which shows no helicity.  Fig.~2a shows Im$(\eta(r))$
calculated by the MDA. As can be seen, it is compatible with
0. In Fig.~3c we plot the results of the algorithm for Re$(\eta(r))$ for
the two fixed initial configurations, showing the independence of the
evolution from the initial state.

Next, we introduce our Multisite Microcanonical method, which is a 
non-local, non-deterministic algorithm allowing the exchange of energy among
different sites, keeping the system energy constant.

Using the notation introduced in (\ref{POLAR}) we rewrite the system
energy as
\begin{equation}
E= \frac{1}{2}{\rm Re} 
   \sum_{n,\pm\mu}\phi(n)\phi^*(n+\mu)
= \frac{1}{2}\sum_{n}\rho(n)\cos (\theta(n)+\alpha(n))
\equiv\frac{1}{2}(\vec \rho , \vec x) \ . 
\end{equation}

Keeping a constant energy means keeping constant the scalar product of the 
two vectors $\vec \rho$ and $\vec x$ just defined: 

\begin{equation}
(\vec \rho , \vec x)=K
\label{SCALAR}
\end{equation}

The expression is valid for the energy of a subset of any $N$ sites, the 
vectors $\vec \rho$ and $\vec x$ being now $N$-dimensional. If, moreover,
the set does not contain interacting sites, updating the set means changing
the vector $\vec x$, as $\vec \rho$ and the $\alpha^\prime$s remain constant 
in the process.

The algorithm proceeds as follows: $N$ non-interacting sites are chosen,
$\vec \rho$ and $\vec x$ are computed, a new $\vec x^\prime$ is generated in
such a way that 
$(\vec \rho , \vec x^\prime)=(\vec \rho , \vec x)$ and the new 
variables $\theta^\prime$ are obtained from $\vec x^\prime$.

Yet, many technical details must be kept in mind. The components of the vector
$\vec x^\prime$ being cosines, must lie in the interval $[-1,1]$, and, when 
computing the new values $\theta^\prime$, one has the freedom to choose the
sign of $\sin (\theta^\prime+\alpha)$, maintaining detailed balance. One 
possibility is choosing it with equal probability for both signs, another is
to include a measure of overrelaxation, choosing the sign which yields the 
$\theta^\prime$ which lies farther away from the original one.

But, in order to ensure ergodicity
and balance, it is the $\phi^\prime$ space (or equivalently, the
 $\theta^\prime$ space) which must be filled uniformly. Once we have generated
the $\vec x^\prime$s uniformly, we can get a uniform distribution in the 
$\theta^\prime$ space weighting the configurations with the Jacobian of the
transformation from the $\theta^\prime$s to the $\vec x^\prime$s, which is
\begin{equation}
J(\theta \to x)=\prod_n \frac{1}{\sin (\theta(n)+\alpha(n))}
\end{equation}
so that, once we have generated uniformly the new $\vec x^\prime$ we keep it
or not with a probability proportional to the quotient of the new and old 
jacobians.

Our problem reduces, then, to uniformly generating points in a 
$(N-1)$-dimensional polyhedron, which is the intersection of the 
$N$-dimensional solid hypercube of 
side two centred in the origin and the $(N-1)$-dimensional hyperplane 
perpendicular to $\vec \rho$ containing $\vec x$.

An obvious way to generate $\vec x$ uniformly is generating uniformly $(N-1)$
components in the interval $[1,-1]$, and finding the $N$th component which
ensures that $\vec x$ lies in the hyperplane. In case the last component lies
in the interval $[1,-1]$, one accepts the new point, otherwise one must start
and try again. Clearly, the inefficiency of this method grows exponentially
with $N$, making the method useless for other than a few sites.

The largest set of non-interacting sites is the set of all equally
coloured sites in the checkerboard lattice, which contains $N=V/2$
points, which for an interesting lattice is of the order of thousands,
which makes the use of such inefficient methods hopeless.

We have explored other more sofisticated methods, on which we shall report
elsewhere, but in general one must trade acceptance for computational load,
the net result being the unimplementability of the method for large $N$.
We have relaxed the demand to generate uniformly at each step, still 
insisting on keeping
detailed balance, but then the explored region becomes small, and the
decorrelation is poor.

Turning to MM with small $N$, $N = 1$ is local and corresponds to
LMO.  $N = 2$ is the smallest value which allows a non-local update,
and consequently an energy exchange between arbitrarily different
sites (LMO exchanges energy only between the ends of a link). That,
together with the fact that at $N = 2$ the efficiency of the method is
highest, the geometrical acceptance rate being one, as will be shown
later, has moved us to choose $N = 2$.

Having chosen the value of $N$, the problem remains of how to pick the
two non-interacting sites. The most unbiased choice is picking them at
random (RMM), and that has been done. Yet, we have mentioned that
sequential updating has the advantage of a wave effect, which
transmits the decorrelation effect along the sequence, so we have
basically done simulations moving one site sequentially and picking
the other one at random, which seems to be efficient.

\begin{figure}
\begin{center}
\epsfig{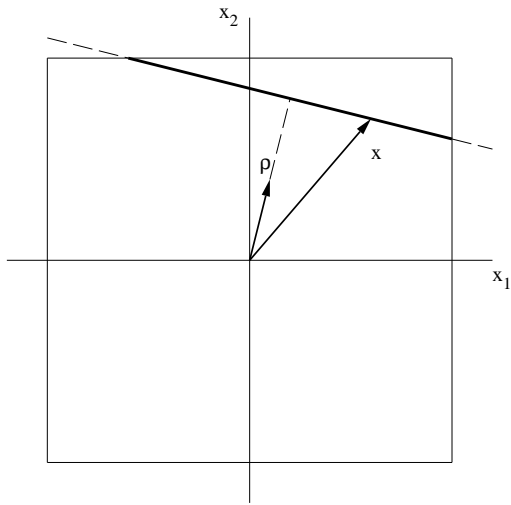}
\caption{Intersection of the solid hypercube of side 2 centered in the 
origin with the hyperplane perpendicular to $\vec \rho$ containing
$\vec x$ in two dimensions.}
\end{center}
\end{figure}

With $N = 2$, the equation
\begin{equation}
\rho_1 x_1 + \rho_2 x_2 = K
\label{SCALAR2}
\end{equation}
must be satisfied, $x_1$ and $x_2$ being uniformly distributed in the allowed 
region. That is easily done; one looks for the maximum and minimum 
values of $x_1$ for which a solution for $x_2$ to (\ref{SCALAR2}) exists 
between $-1$ and $1$, and then a value for $x_1$ is generated randomly and 
uniformly in that interval. Fig.~4 illustrates the process.

Let us call $(x_1, x_2)$ the old values of $\vec x$ and 
$({x^\prime}_1 ,{x^\prime}_2)$ the new ones, generated as described above.
From them the values of the new 
$({\theta^\prime}_1 , {\theta^\prime}_2 )$ are obtained,
solving $x^\prime=\cos (\theta^\prime + \alpha )$. Next, in order to obtain
uniformly distributed $\theta^\prime$s, the quotient $C$ of jacobians is 
computed
\begin{equation}
C=\frac{{(\sin ({\theta^\prime}_1+{\alpha^\prime}_1)
	  \sin ({\theta^\prime}_2+{\alpha^\prime}_2))}^{-1}} 
       {{(\sin ({\theta}_1+{\alpha}_1)
	  \sin ({\theta}_2+{\alpha}_2))}^{-1}} 
\end{equation}
and the new values are accepted if $C \ge 1$, or otherwise with probability 
proportional to C. The acceptance from this factor becomes around 70\%.
Finally we have chosen to take the sign of 
$\sin (\theta^\prime + \alpha )$ which yields the farthest $\theta^\prime$ 
from the original one.

The results of the simulation with the implementation of the MM algorithm
just described, are as follows: Fig.~1d shows the evolution of the 
magnetization phase, which rambles unsystematically.
Fig.~2b shows that Im$(\eta(r))$ is compatible with $0$, with much smaller 
errors than for other algorithms. Fig.~3d , in which results of RMM runs
are included, shows that Re$(\eta(r))$
behaves identically for both initial configurations and both implementations
of the algorithm, with smaller errors for the choice sequential-random.

MM results for Re$(\eta(r))$ are just outside error bars with respect to MDA
results. This is due to the fact, already mentioned, that MDA simulations
are not strictly microcanonical, and in fact the energy values of the
MDA simulation, which run in the range $(0.4550 , 0.4562)$, yield a slightly
higher mean value than those of the MM simulations, which range in
$(0.4552 , 0.4553)$ (due to rounding). 
This explains qualitatively why the MDA results are 
slightly bigger than the MM ones. This situation is unavoidable, if we
insist on studying the dependence of the results on the initial state, which 
forces us to use fixed initial states for all our simulations, so losing in
a certain measure our control on the average energy in the case of MDA.

Forgetting about helicity, we have run with MM at energy $0.4556$ (the mean 
energy of MDA) and 
in this case the results are compatible.

The comparison with the LMO results in Fig.~3a,b is pointless, since
those results are selfincompatible.

Our conclusions are, then:

SLMO is not ergodic (for the two-dimensional XY model).

We have introduced a new multisite, microcanonical update method, MM.

Among the microcanonical algorithms studied here, SLMO and MM have proved to 
be the most efficient, as far as Monte Carlo time decorrelation is concerned,
although LMO runs only over a restricted region of phase space.

The RLMO, MDA and MM algorithms yield compatible results, while LMO
results are start-dependent and incompatible. 

\bigskip

We wish to thank Juan J. Ruiz-Lorenzo for useful discussions.
Partially supported
by CICyT AEN93-0604-C03, AEN94-0218, AEN95-1284E, and ERBCHRXCT920051.
D.I.
is a MEC Fellow.

\end{document}